# Two-phase Cryogenic Avalanche Detectors with THGEM and hybrid THGEM/GEM multipliers operated in Ar and Ar+N$_2$

A. Bondar,[a,b] A. Buzulutskov,[a,b,*] A. Dolgov,[b] A. Grebenuk,[a] E. Shemyakina,[a,b] A. Sokolov,[a,b] D. Akimov,[c] A. Breskin[d] and D. Thers[e]

[a] *Budker Institute of Nuclear Physics SB RAS, Lavrentiev avenue 11, 630090 Novosibirsk, Russia*
[b] *Novosibirsk State University, Pirogov street 2, 630090 Novosibirsk, Russia*
[c] *Institute for Theoretical and Experimental Physics, Bolshaya Cheremushkinskaya 25, 117218 Moscow, Russia*
[d] *Weizmann Institute of Science, 76100 Rehovot, Israel*
[e] *Laboratoire SUBATECH, UMR 6457 Ecole des Mines, CNRS/IN2P3 and Universite de Nantes, 4 rue Alfred Kastler, 44307 Nantes Cedex 3, France*

E-mail: `A.F.Buzulutskov@inp.nsk.su`

ABSTRACT: Two-phase Cryogenic Avalanche Detectors (CRADs) with GEM and THGEM multipliers have become an emerging potential technique for charge recording in rare-event experiments. In this work we present the performance of two-phase CRADs operated in Ar and Ar+N$_2$. Detectors with sensitive area of $10\times10$ cm$^2$, reaching a litre-scale active volume, yielded gains of the order of 1000 with a double-THGEM multiplier. Higher gains, of about 5000, have been attained in two-phase Ar CRADs with a hybrid triple-stage multiplier, comprising of a double-THGEM followed by a GEM. The performance of two-phase CRADs in Ar doped with N$_2$ (0.1-0.6%) yielded faster signals and similar gains compared to the operation in two-phase Ar. The applicability to rare-event experiments is discussed.

---

[*] Corresponding author.

**Contents**

# 1. Introduction

In recent years, two-phase Cryogenic Avalanche Detectors (CRADs) have become an emerging technique proposed for rare-event experiments such as dark matter search, coherent neutrino-nucleus scattering and astrophysics neutrino detection (see reviews [1], [2]). In two-phase CRADs the primary ionization charge, composed of electrons produced in the noble liquid and emitted into the gas phase, is multiplied in the gas phase with GEM [3] or thick GEM (THGEM) multipliers [4].

In terms of the maximum gain, most promising results have been obtained with two-phase CRADs operated in Ar: charge gains as high as $10^4$ and $3\times10^3$ with the triple-GEM [1],[5],[6] and double-THGEM [1],[7] multipliers respectively were routinely attained at temperatures of 84-87 K in multipliers having 3×3 cm$^2$ active area. A gain of about 80, was obtained in a two-phase Ar CRAD having a 2D readout, with single-THGEM multiplier of a larger 10×10 cm$^2$ active area [8]. Here the CRAD charge gain is defined as the ratio of the output anode charge of the multiplier to the input "primary" charge, i.e. that prior to multiplication.

However, several problems in two-phase CRAD performances remain unsolved. The first is that of the gain limit. Ultimate sensitivity, i.e. minimum threshold, requires operation in single-electron counting mode, in a 2D readout and self-triggering mode and with a minimum noise level. For GEM- and THGEM-based two-phase CRADs this requires stable operation at gains of about 20,000-30,000 [9], while today's limit is about 4-8 times lower. The so far achieved level corresponds to the detection threshold of 4 primary electrons with a triple-GEM and 20 electrons with a double-THGEM [7], at gains of 5000 and 1500 respectively reached with multiplier active area of about 3×3 cm$^2$ at temperatures of 84-87 K. The gain limit in larger, practical-size multipliers could be lower due to the increased number of holes.

For effective operation in single-electron counting mode, the obvious way to increase the gain would be an increase in the number of stages, until 4-5 stages in the case of GEMs and 3-4 stages in the case of THGEMs. In the latter case it may be technically difficult due to higher total operating voltages. Therefore, it looks attractive to combine different types of MPGD multipliers, similarly to that studied in [10],[11] where THGEMs were combined with PIM and Micromegas multipliers. In these works gains of $6\times10^6$ were obtained in Ne+10%$CF_4$ with X-rays, at 171 K, in the hybrid THGEM/PIM/Micromegas detector; it was used in a gaseous photomultiplier (GPM) recording primary scintillation light in a liquid Xe TPC. In the current work we present another configuration of a hybrid multiplier, combining thin and thick GEMs.

Another way to increase the CRAD gain limit was proposed in [12]: it was suggested to add a $N_2$ additive to two-phase Ar. This was based on the observation that the GEM or THGEM maximum gain in mixtures of Ar and $N_2$, even with minute $N_2$ contaminations, can be considerably higher compared to that of pure Ar [13],[14]. This will be further examined in the present work.

It should be mentioned that there exist an alternative way to increase the overall CRAD gain, namely by optically recording THGEM signals with Geiger-mode APDs (GAPDs) [15],[16],[17],[18] or Large-Area APDs (LAAPDs) [19]. In this case, the THGEM gain limit can be substantially reduced, compensated by the high GAPD gain. The reader is referred to a recent review [1].

The second problem in CRAD performances is that of the resistance to electrical discharges of "standard" (thin Kapton) GEMs. In our laboratory, when operating two-phase CRADs at maximum gains (approaching 10,000), it was observed that triple-GEMs were not able to withstand electrical discharges on a long term: after several series of measurements the GEM's maximum reachable gain decreased by several times. Obviously, this is due the low resistance of thin GEMs to discharges, as a result of metal evaporation from the electrodes and its deposition on the insulator in the GEM holes. The solution of the problem might be switching to thicker THGEMs that were found to behave in more reliable way under discharges, or else combining THGEM and GEM multipliers.

The third problem is that of natural radioactivity of standard G10 (or FR4) material of THGEMs. These contain glass fibres, with radioactive $^{40}$K as main source of undesirable background. To reduce background in rare-event experiments, other, radio-clean materials should be investigated, in particular polyimide (e.g. Kapton or Cirlex) [20].

In this work we address these problems, presenting some new features in two-phase Ar CRAD design and studying their effect on CRADs performances. It should be emphasized that these studies were performed here for two-phase Ar CRADs having a practical active volume, namely $10\times10$ cm$^2$ active area and 1-5 cm thick liquid layer. Accordingly, the new features of the two-phase CRADs studied here include:

- Combining a double-THGEM and a GEM in a hybrid triple-stage multiplier;
- Doping two-phase Ar with $N_2$;
- Evaluating polyimide-made double-THGEM detectors.

## 2. Experimental setup

The experimental setup was similar to that used in our previous measurements with two-phase Ar and Xe detectors [16],[21]. It includes a 9 l volume cryogenic chamber of 24 cm internal diameter, cooled with liquid nitrogen. The chamber comprised of a double-THGEM or double-THGEM/GEM placed within the saturated vapour above the liquid; the cathode electrode was placed at the bottom, immersed in either a 1 or 5 cm thick liquid Ar layer (see Fig. 1). The corresponding active volume of liquid Ar was either 0.1 or 0.5 l (thus of the order of a cubic decimetre in the latter case), with a total amount of liquid Ar of either 0.5 or 2.4 l respectively. The detector was operated in two-phase Ar in equilibrium state, at a saturated vapour pressure of 1.0 atm corresponding to a temperature of 87 K.

After each mounting stage, the chamber was evacuated with a turbo-molecular pump to a vacuum level of $2\times 10^{-5}$ Torr, and baked at 50 °C for several tens of hours. In case of two-phase CRAD operated in Ar, the gas was taken from the bottle with a specified purity of 99.998% ($N_2$ content <10 ppm); during cooling procedures it was additionally purified from oxygen and water by Oxisorb filter [22], providing electron life-time in the liquid of 13 µs (see section 3.2 in the following), which corresponds to an oxygen-equivalent impurity content of 30 ppb.

In case of two-phase CRAD operated in Ar+$N_2$, the $N_2$ content in the liquid and gas phase was calculated similarly to that of [12], i.e. using "Raoult" law [23]:

$$P(N_2\ in\ gas) = P(N_2\ saturated) \cdot X(N_2\ in\ liquid)$$

Here $P(N_2\ in\ gas)$ is the $N_2$ partial pressure in the gas phase; $P(N_2\ saturated)$ is the saturated $N_2$ vapour pressure at a given temperature which is equal to 2.73 atm at 87 K [24]; $X(N_2\ in\ liquid)$ is the $N_2$ content in the liquid. This dependence is well reproduced by the experimental $N_2$ content diagram in two-phase Ar+$N_2$ system presented elsewhere [25]. In this work the measurements were performed at two values of $N_2$ content in the liquid, namely at 0.11% and 0.55% corresponding to $N_2$ content in the gas phase of 0.30% and 1.5% respectively, at 87 K. To prepare such Ar+$N_2$ mixtures, the required amount of $N_2$ was calculated and doped into the Ar gas bottle before the cooling procedure.

The liquid level in the gap underneath the first THGEM (THGEM1), namely in the cathode gap in Fig. 1 (top, middle) and interface gap in Fig. 1 (bottom), was monitored with an accuracy of 0.5 mm using two methods. Firstly, the interface or cathode gap capacitance, unambiguously related to the liquid layer thickness, was measured during liquefaction procedure, similarly to that of [6],[21]. Secondly, the liquid layer thickness was calculated from the amount of condensed Ar using the well-defined geometry of the chamber bottom. Both methods yielded similar results.

Three types of THGEM and GEM plates of a decimetric size were used in the measurements; their geometrical parameters are presented in Table 1. The first two types were manufactured by the CERN PCB workshop; these were a THGEM made of G10, sometimes denoted in the following as THGEM(G10), and a "standard" (thin) GEM made of Kapton, both types having an active area of $10\times 10$ cm$^2$. The third type was a THGEM made of polyimide, a material similar to Kapton, manufactured by CPTA company (Russia) [26]; it was somewhat thinner compared to the THGEM(G10) and had smaller holes.

Since the majority of the results were obtained with THGEM(G10), in the following data presentation it will be implied that the THGEMs are made of G10, unless otherwise specified.

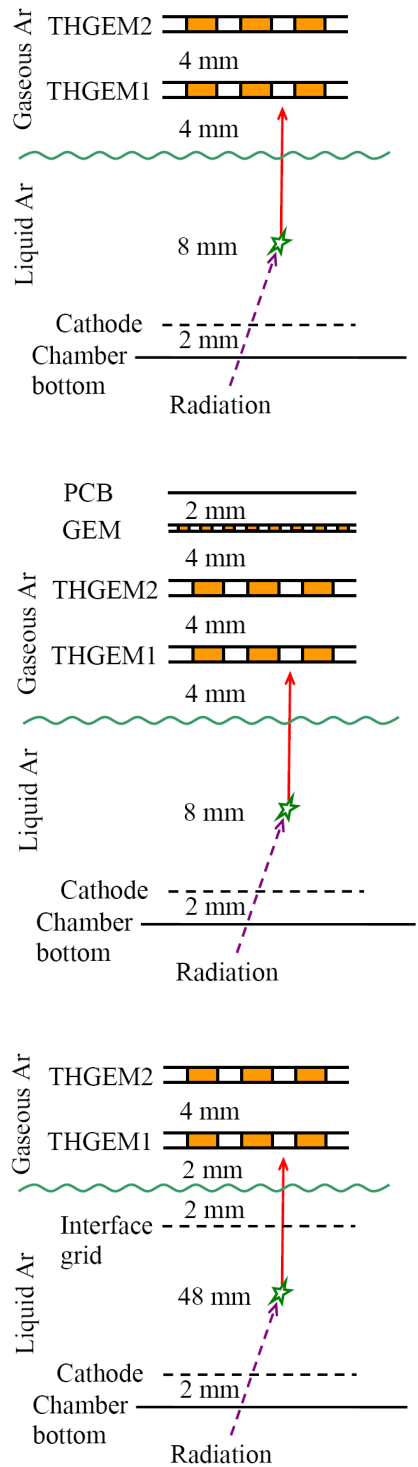

Fig. 1. Schematic views of two-phase Ar CRADs of different types used in this work (not to scale): with a G10 or polyimide double-THGEM multiplier and 1 cm thick liquid-argon layer (top), a hybrid double-THGEM/GEM/PCB multiplier and 1 cm thick liquid-argon layer (middle) and a double-THGEM multiplier and 5 cm thick liquid-argon layer (bottom).

| Type (dielectric material) | THGEM (G10) | GEM (Kapton) | THGEM (polyimide) |
|---|---|---|---|
| Manufacturer | CERN workshop | CERN workshop | CPTA |
| Active area | 10×10 cm$^2$ | 10×10 cm$^2$ | 5 cm in diameter |
| Dielectric thickness (t), mm | 0.4 | 0.05 | 0.25 |
| Hole pitch (p), mm | 0.9 | 0.14 | 0.7 |
| Hole diameter in dielectric (d), mm | 0.5 | - | 0.3 |
| Hole rim (h), mm | 0.1 | (double-conical) | 0.1 |
| Hole diameter in electrodes, mm | 0.7 | 0.075 | 0.5 |

Table 1. Geometrical parameters of decimetric THGEM and GEM electrodes used in this work. The copper layer thickness on the THGEM electrodes was typically 30 μm.

In one particular measurement, THGEM(G10) electrodes of a fourth type, similar to that of our previous works [7],[16],[21], were used. These were produced by Print Electronics (Israel) and had the same geometrical parameters as that of the THGEM(G10) of Table 1, except of the active area which was smaller, of 2.5×2.5 cm$^2$.

To study new CRAD features discussed in the introduction, several sets of measurements were conducted using four different CRAD designs, listed below in a chronological order:

- Two-phase CRAD with a double-THGEM multiplier, active area of 2.5×2.5 cm$^2$ and 1 cm thick liquid layer (see Fig. 1, top).
- Two-phase CRAD with a hybrid triple-stage multiplier, namely a double-THGEM/GEM/PCB multiplier, active area of 10×10 cm$^2$ and 1 cm thick liquid layer (see Fig. 1, middle). Here PCB defines a Printed Circuit Board anode; in this work it was a passive THGEM electrode.
- Two-phase CRAD with a double-THGEM multiplier, active area of 10×10 cm$^2$ and 5 cm thick liquid layer (see Fig. 1, bottom). The characteristic feature of this design was an interface grid located just below the liquid surface, to permit setting different electric fields within the cathode and interface gaps (between the cathode and interface grid and between the interface grid and first THGEM, respectively; Fig. 1, bottom). The interface grid was made of 100 μm diameter wires having 1 mm spacing.
- Two-phase CRAD with a polyimide double-THGEM multiplier, active area of 5 cm in diameter and 1 cm thick liquid layer (see Fig. 1, top).

In all designs, the multiplier assembly was terminated from the top side by a support G10 plate (not shown in Fig. 1). Figs. 2 and 3 show photographs of the double-THGEM(G10)/GEM/PCB assembly at the mounting stage and that of the polyimide double-THGEM.

The cathode, interface grid (if any) and multiplier electrodes were biased through a high-voltage resistive divider, placed outside the cryostat. The dividers for the CRAD with the double-THGEM/GEM/PCB multiplier operated in a single-THGEM, double-THGEM and double-THGEM/GEM/PCB modes are schematically depicted in Fig. 4. Following our traditional designations [1],[7], these configurations are denoted in the following as "1THGEM", "2THGEM" and "2THGEM/GEM/PCB" respectively. In these modes the anode signals were recorded from the last electrode of the first or second THGEM and from the PCB, respectively (see Fig. 4). In the 1THGEM and 2THGEM modes the other electrodes (not connected to the divider) were kept floating; this guaranteed that the field-lines are terminated at the readout electrode, providing full collection of the avalanche electrons.

For the two-phase CRADs with double-THGEM multiplier assembly (G10 or polyimide), the divider network was similar to that of the 2THGEM in Fig. 4.

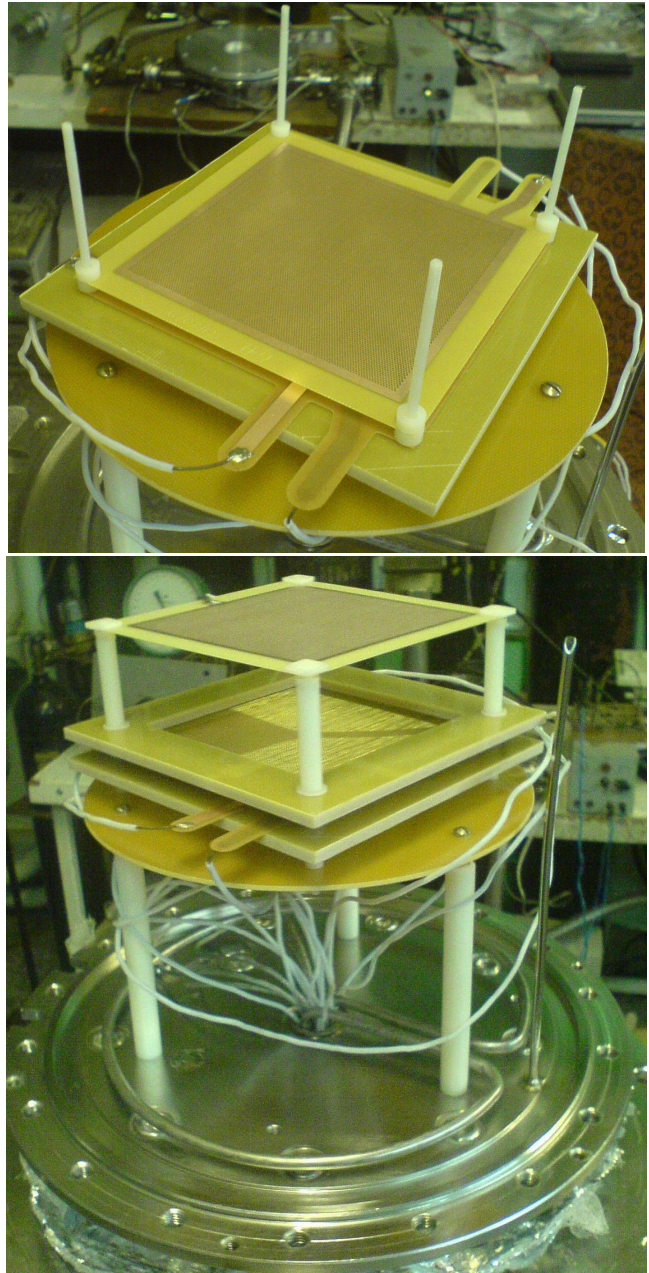

Fig. 2. Photographs of the double-THGEM(G10) assembly (top) and cathode/interface-grid/double-THGEM assembly (bottom) at the mounting stage of the two-phase Ar CRAD with the double-THGEM multiplier; 10×10 cm$^2$ active area; 5 cm thick liquid layer. Here the cathode is a passive THGEM plate.

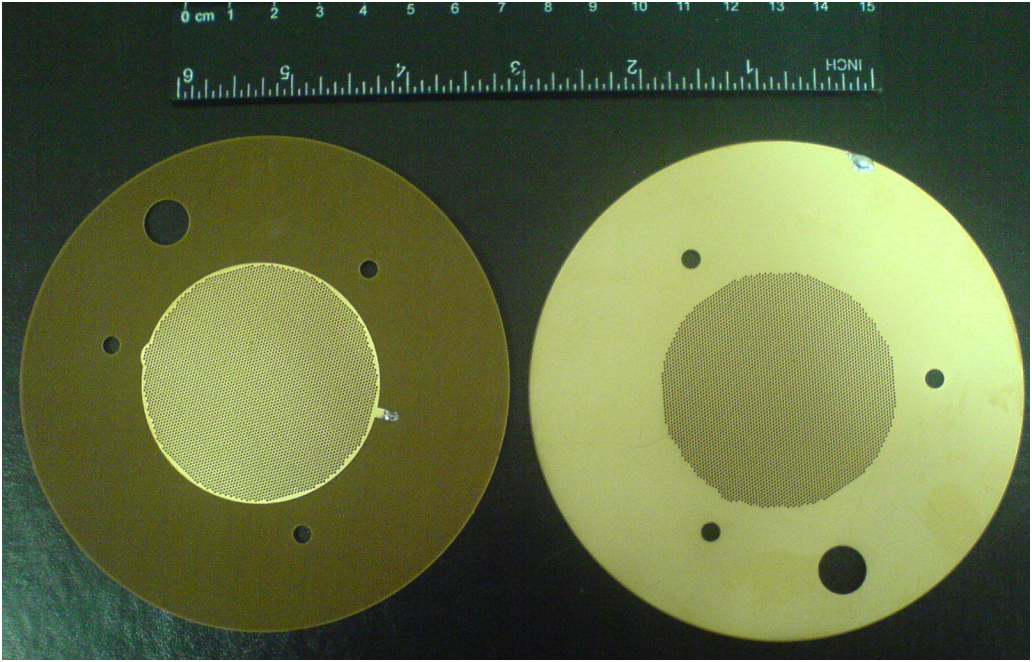

Fig. 3. Photograph of the polyimide THGEMs.

The radiation-induced avalanche (charge) signal was read out from the last-multiplier electrode in a given readout mode using a charge-sensitive amplifier with a shaping time of either 0.5 or 10 μs and sensitivity in the latter case of about 10 V/pC. The amplifier was kept outside the cryogenic chamber.

It should be remarked that in the 2THGEM/GEM/PCB readout mode the charge gain typically amounted to 0.3-0.5 of that of the direct connection to the GEM electrode (i.e. of 2THGEM/GEM mode), since the avalanche charge was not fully transferred to the PCB from the GEM holes [27], a fraction being collected at the GEM electrode. On the other hand, the PCB readout mode is usually required for 1D or 2D position sensitivity recording by means of a patterned readout anode.

Note that in the 2THGEM and 1THGEM modes the first electrode of the readout THGEM was grounded through a rather large capacitance (C1 in Fig. 4), its value substantially exceeding the THGEM's capacitance (~1000 pF). Its role was to ground at high-frequencies the (positive) signal induced by the movement of avalanche ions through the THGEM holes; otherwise this signal would have been induced on the readout electrode through capacitive coupling, thus neutralizing the useful (negative) signal.

It should be also remarked that the rather large GEM capacitance (~6000 pF) prevented us to realize the 2THGEM/GEM readout mode (namely, without a PCB anode) due to the fact that the charge amplifier was not able to properly operate with such a large input capacitance.

The signals in the detector were induced by either X-rays from a $^{241}$Am source having among others a 60 keV line (here of few tens Hz) or 15-40 keV X-rays from a pulsed tube, with a pulse-width of 0.5 μs and frequency of 240 Hz [28]. The detector was irradiated from outside, practically uniformly across the active area, through two 1 mm thick and 5 cm diameter Al windows located at the chamber's bottom.

.

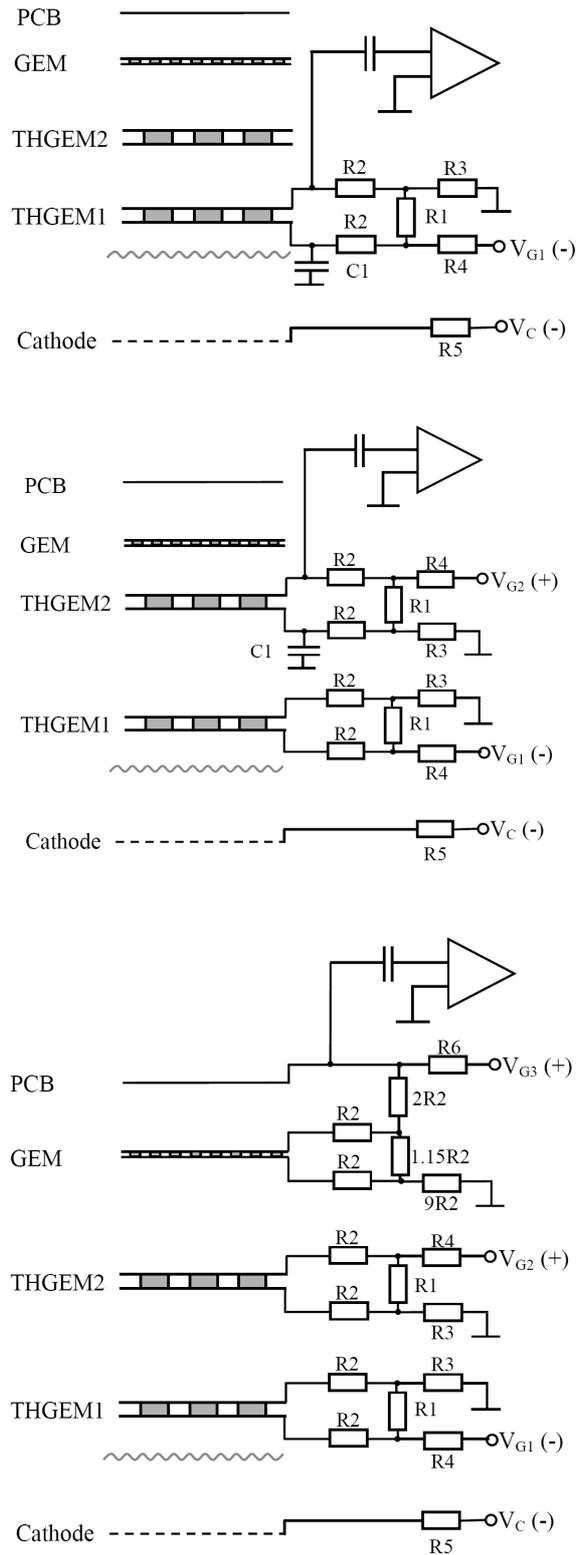

Fig. 4. Schematic view of the high voltage divider of the two-phase CRAD with a double-THGEM/GEM/PCB multiplier operated in 1THGEM (top), 2THGEM (middle) and 2THGEM/GEM/PCB (bottom) modes. R1=8.6 MΩ; R2=68 MΩ; R3=2.15 MΩ; R4=1 MΩ; R5=4.3 MΩ; R6=3.3 MΩ; C1=9400 pF.

The CRAD's charge gain was measured with pulsed X-rays with an amplifier shaping time of 10 μs, similar to that in our previous works [6],[7],[21]: the charge gain is defined as the pulse-height of the avalanche signal of the THGEM or THGEM/GEM/PCB multiplier, recorded at its anode ( i.e. at the top electrode of the last THGEM for the former and at the PCB for the latter), divided by that of a calibration signal. The latter was recorded at the first electrode of the first THGEM, the cathode or interface gap being operated in an ionization collection mode at corresponding electric field. At higher gains, the gain-voltage dependence was measured using the relative peak position in the pulse-height spectrum, induced by 60 keV X-rays. The two sets of measurements were merged, providing the absolute gain values. The maximum gain was defined at the onset of occasional discharges. The reproducibility of the gain value at a given voltage in the two-phase mode was about 10%.

Other details of the experimental setup and procedures are presented elsewhere [16],[21].

## 3. Two-phase CRADs with 2THGEM and 2THGEM/GEM/PCB multipliers in Ar

### 3.1 Active area: 10×10 cm$^2$ versus 2.5×2.5 cm$^2$ with 1 cm liquid layer

Fig. 5 illustrates the gain characteristics of the two-phase Ar CRAD with the double-THGEM multiplier, having an active area of 10×10 cm$^2$ and 1 cm thick liquid Ar layer; it is shown in comparison with that of a 2.5×2.5 cm$^2$ active area detector.

One can see that both characteristics are rather similar, indicating upon the proper performance of the two-phase Ar CRAD with a decimetric active area. On the other hand, the maximum gain of 1200 recorded with the 10×10 cm$^2$ detector, limited here by discharges, is 3-fold lower, compared to that reported elsewhere for the 2.5×2.5 cm$^2$ one [7]. The difference could result from the larger number of holes, implying a larger discharge probability on defects.

This maximum gain value, of the order of 1000, obtained here for the 10×10 cm$^2$ two-phase Ar CRAD, is obviously not sufficient for charge readout in a single-electron counting mode. Nevertheless, it is still significant and might be sufficient for other applications; in particular, the modest charge gain could be overcome by applying GAPD-based or other optical readout concepts [1],[15],[16],[17],[18],[19].

Fig. 6 (top) illustrates the pulse-height distributions induced by X-rays from the $^{241}$Am source, of the two-phase Ar CRAD with 10×10 cm$^2$ active area; anode signals from the double-THGEM were recorded at a gain of 460. This distribution is rather similar to that previously recorded from ~3×3 cm$^2$ two-phase Ar CRADs with a double-THGEM and 1 cm thick liquid Ar layer [7],[16] and others recorded with a triple-GEM multiplier [6]; note that in all these works the 60 keV X-ray peak had a symmetrical shape and its width corresponded to an amplitude resolution of σ/A~20%. That means that the amplitude resolution of the two-phase Ar CRAD does not depend on gain non-uniformity over the active area, but rather on other physical factors.

### 3.2 CRAD with 5 cm liquid layer

In this paragraph we consider a two-phase Ar CRAD with a THGEM multiplier coupled to a 5 cm thick liquid layer. Before studying the gain characteristics, the electron life-time in liquid Ar was estimated, analyzing the shape of the ionization signal in liquid Ar within the cathode gap. Fig. 7 shows a signal induced by pulsed X-rays, at a drift field of 1.25 kV/cm; it was recorded at the interface grid, the cathode gap being operated in an ionization collection mode, i.e. with no high voltage applied across the interface gap. After traversing the 2 mm thick Al windows at the chamber bottom, the X-ray spectrum consists of a 17 keV line (63%) and a continuous distribution with a maximum at 20 keV (37%) [28]. Such soft X-rays are mostly converted close to the cathode, resulting in a trapezoidal shape of the anode signal, clearly seen in Fig. 7. The total signal width (at the base), of 22 μs, corresponds to the electron drift time between the cathode and the interface grid, taking into account the electron drift velocity in liquid Ar of 2.2 mm/μs at 1.25 kV/cm [29].

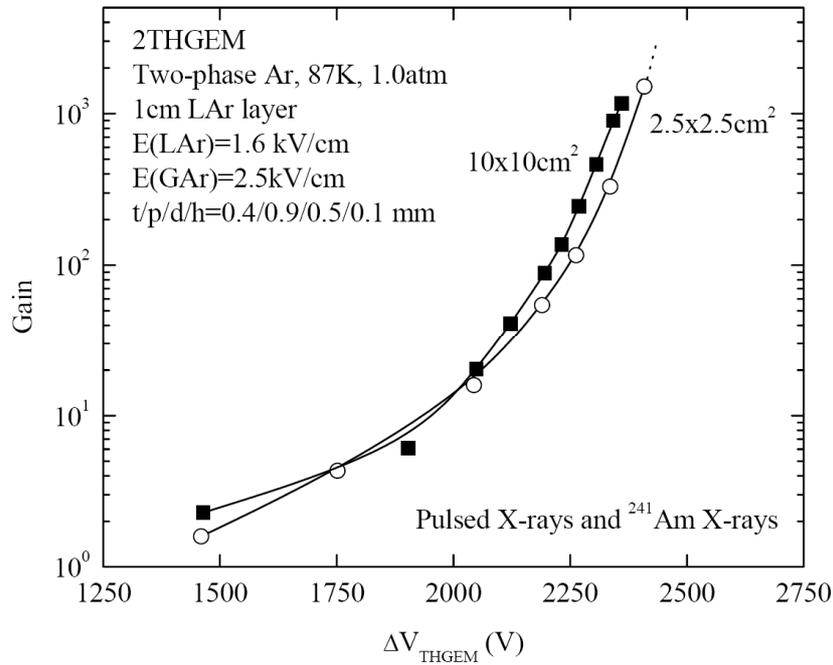

Fig. 5. Gain-voltage characteristics in the two-phase Ar CRAD with the 2THGEM multiplier and 1 cm thick liquid layer, having a 10×10 cm$^2$ active area; here the maximum gain was limited by discharges. The gain curve of a 2.5×2.5 cm$^2$ active area detector is shown for comparison, for which the discharge limit was not attained. The electric fields in the liquid and gas phases at the liquid-gas interface and the THGEM parameters are indicated in the figure.

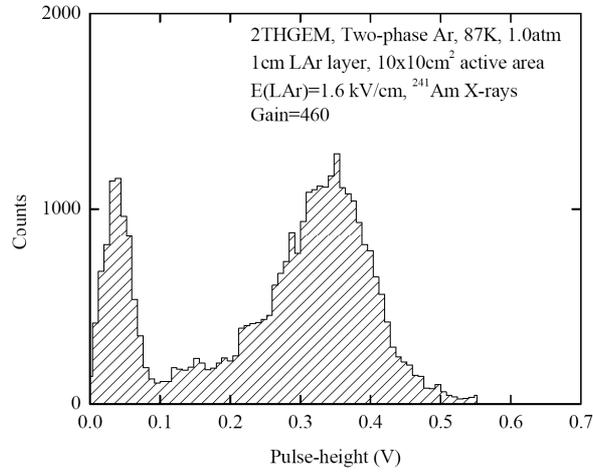

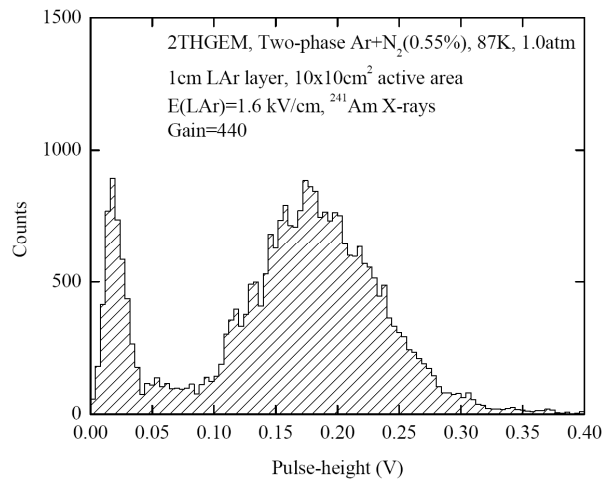

Fig. 6. Pulse-height distributions of anode signals in the 10×10 cm$^2$ two-phase CRAD with the 2THGEM multiplier and 1 cm thick liquid layer, operated in Ar (top) and Ar+N$_2$ (0.55% in the liquid, 1.5% in the gas) (bottom), at respective gains of 460 and 440. The signals were induced by X-rays from $^{241}$Am source. Electric field across the liquid: 1.6 kV/cm.

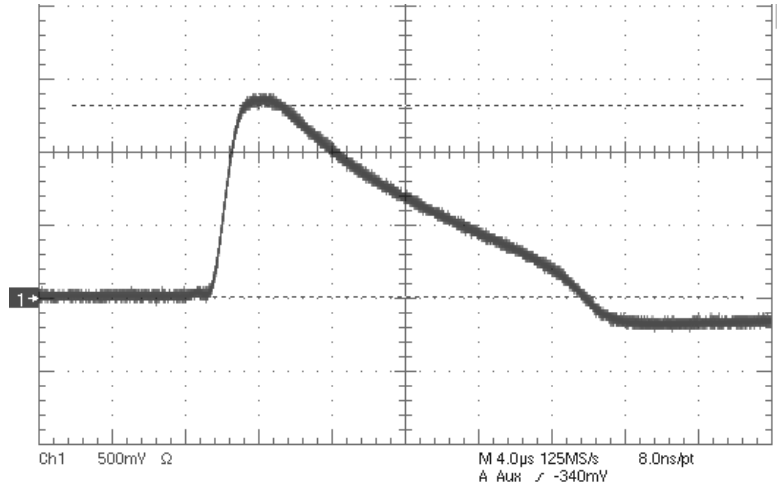

Fig. 7. Collected-charge signal (no multiplication) in liquid Ar induced by pulsed X-rays converted in the 4.8 cm thick cathode gap, under a drift field of 1.25 kV/cm. The signal was recorded at the interface grid of the two-phase Ar CRAD of Fig.1 (bottom). The charge amplifier shaping time is 0.5 μs; the time scale is 4 μs/div; the vertical scale is 500 mV/div.

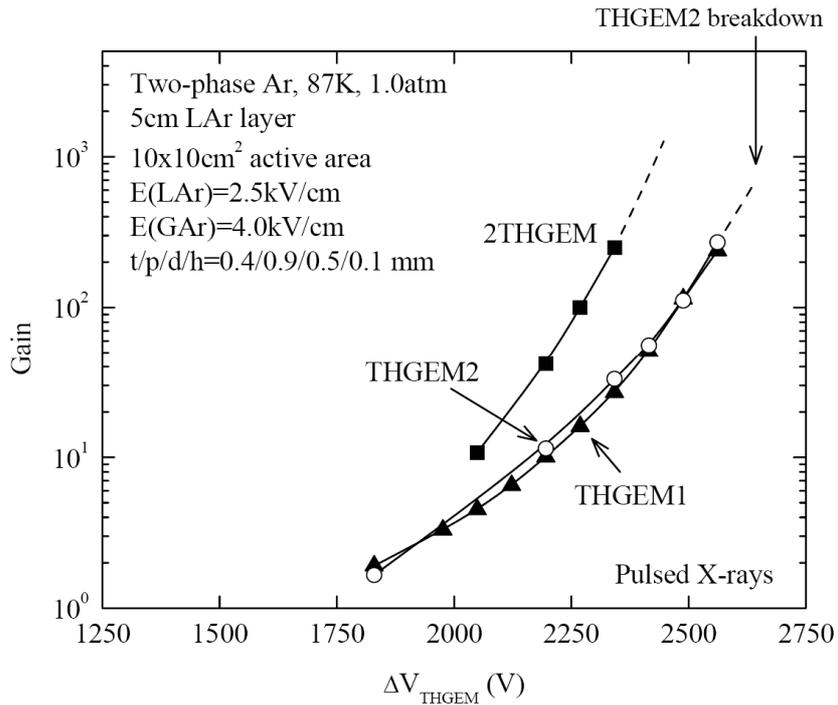

Fig. 8. Gain-voltage characteristics in the two-phase Ar CRAD with 10×10 cm$^2$ 1THGEM and 2THGEM multiplier configurations and 5 cm thick liquid layer. The former is denoted THGEM1; the gain-curve of the second THGEM denoted as THGEM2 is shown as well. Gain limits due to discharges were not reached for the 2THGEM and THGEM1 configurations, while that for THGEM2 was reached at the voltage indicated in the figure. The electric fields in the liquid and gas phases at the liquid-gas interface and THGEM parameters are indicated in the figure. The electric drift field in the cathode gap was 0.63 kV/cm.

The electron life-time in the liquid can be evaluated from this signal shape, subtracting the baseline and fitting the signal with an exponentially decaying function; it amounts at 13 μs. This is 1.5-fold lower compared to our previous results obtained with CRADs of 3×3 cm$^2$ active area [6],[12]; it could be a consequence of impurities resulting from a larger amount of detector materials in the present cryogenic chamber, inducing higher outgassing rates. This problem will be addressed in a forthcoming CRAD design with a new Ar purification and recirculation system.

Fig. 8 shows the gain characteristics of the two-phase Ar CRAD with a 10×10 cm$^2$ active area and 5 cm thick liquid Ar layer, in a 2THGEM and 1THGEM readout mode. Here the gain limits due to discharges were not reached, to preserve the THGEMs for further measurements. One can see that the double-THGEM characteristic for the 5 cm thick liquid layer looks similar to that of the 1 cm one (shown in Fig. 5). In addition, the maximum gain of the single-THGEM reached a rather large value, of about 300. These observations confirm the statement that increasing the amount of condensed Ar (up to a 2.4 l in this case), which typically results in reducing the temperature gradients within the cryogenic chamber, does not affect the performance of the THGEM multiplier in saturated Ar vapour.

On the other hand, we tried to determine the maximum gain limit for the second THGEM in these conditions, i.e. in saturated Ar vapour at 87 K. A gain limit as high as 700 was reached at the maximum voltage at which a considerable breakdown occurred, leading to a partial THGEM damage (see Fig. 8): after the breakdown, the gain limit of the second THGEM decreased down to a few hundreds.

Finally, we confirmed here the proper performance of the two-phase CRAD with the THGEM multiplier having an active volume, of the order of a litre, with relatively high charge gains approaching 1000.

### 3.3 2THGEM/GEM/PCB versus 2THGEM

Combining thick and thin GEMs takes obvious advantage of combining reliability of the former and lower operating voltage of the latter. In addition, combining the THGEM and GEM multipliers in that order, with the THGEM followed by a GEM, is based on the idea that the avalanche charge from a THGEM hole would be distributed between a number of GEM holes; this would reduce the final avalanche charge density, believed to be responsible for the discharge development, and thus increase the discharge limit.

Fig. 9 illustrates the gain characteristics of the two-phase Ar CRAD with the triple-stage 2THGEM/GEM/PCB multiplier, having a 10×10 cm$^2$ active area and 1 cm thick liquid Ar layer. One can see that the overall gain in this configuration reaches a value of 4500, namely about 5-fold larger than that of the 2THGEM (compare to Fig. 5). Note that even for a moderate gain of the double-THGEM substructure, of 240, the overall gain of the 2THGEM/GEM/PCB multiplier exceeds 1000. It should be remarked that these maximum gains in a PCB readout mode have practical importance for position-sensitive charge readout with patterned anodes.

Note that the "real" gain of such structure is indeed 2-3 fold higher due to the avalanche charge sharing between the GEM and the PCB electrodes [27], as discussed in section 2. That means that the overall avalanche gain in GEM holes can be as high as 9,000-14,000, which looks very attractive for applications in rare-event detectors of ultimate sensitivity with optical readout using GAPDs [1],[17] and LAAPDs [19], i.e. in those operated in a single electron counting mode.

On the other hand, the maximum charge gain obtained, of about 5000, is still a factor of 4 lower compared to that required for an effective operation in a single-electron counting mode with charge readout. For the latter, gains as high as 20,000 are needed, taking into account the charge sharing between readout strips or pads. Nevertheless, one may conclude that the hybrid 2THGEM/GEM/PCB multiplier studied here may provide practical charge readout of two-phase Ar CRAD at a rather high sensitivity, which might be still attractive for applications in dark matter search and coherent neutrino-scattering experiments [30].

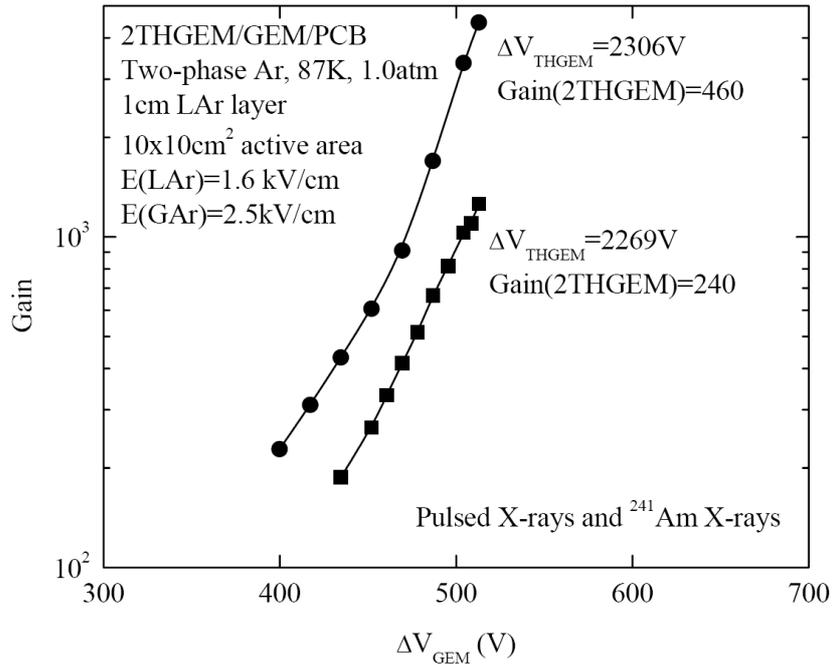

Fig. 9. Gain-voltage characteristics in the 10×10 cm$^2$ two-phase Ar CRAD with the 2THGEM/GEM/PCB multiplier and 1 cm thick liquid layer. Shown is the overall multiplier gain as a function of the voltage across the GEM, at two fixed voltages across each THGEM, i.e. at two values of the 2THGEM gain indicated in the figure. Here the maximum gains were limited by discharges. The electric fields in the liquid and gas phase at the liquid-gas interface are indicated in the figure.

# 4. Two-phase CRADs with 2THGEM and 2THGEM/GEM/PCB multipliers in Ar+$N_2$

The idea of doping liquid Ar with $N_2$ to improve the performance of either liquid Ar TPCs [31],[32],[33] or two-phase Ar CRADs [12] has appeared repeatedly. In the latter case, an interesting effect appeared: the slow component of the electron emission through the liquid-gas interface almost fully converted to the fast emission one [12]. This would make two-phase Ar+$N_2$ CRADs substantially faster compared to that with pure Ar. Indeed, in two-phase Ar CRADs, at typical operating fields of ~2 kV/cm in the liquid, the slow component has a time constant of ~5 μs and amounts to ~60% of the total signal [12], which makes them relatively slow. Note that the observations of the slow component were made possible through the use of fast GEM multipliers [12] or that of THGEM with GAPD-based optical readout [16].

In this work we further confirmed the effect of $N_2$ doping on the slow electron emission component: this is seen in Fig. 10 showing typical anode signals in the two-phase CRADs with the (fast) 2THGEM/GEM/PCB multiplier in Ar and Ar+$N_2$ (0.11% in the liquid, 0.30% in the gas). In Ar, the fast and slow electron emission components are distinctly seen, the slow component being dominant, at an electric field in the liquid of 1.6 kV/cm. In contrast, in Ar+$N_2$ the slow component almost fully disappeared at this electric field.

In addition, the two-phase Ar+$N_2$ CRAD was expected by us to have higher maximum gains [1],[12]. This expectation was based on the observations that doping Ar with $N_2$ at room temperature may reduce photon feedback due to the shift of the energetic Ar emission into the visible spectral range, resulting in an increase of the maximum gain of both the GEM [13] and THGEM [14] multipliers compared to the operation in pure Ar.

Unfortunately in this work we failed to validate this assumption. This is seen from Figs. 11 and 12 showing the gain characteristics in the two-phase Ar+$N_2$ CRADs with the 2THGEM and 2THGEM/GEM/PCB multipliers respectively. One can see that while the operation voltage increased with the $N_2$ content, the maximum gain did not increase as expected; moreover, it decreased (see Fig. 11). In particular in the two-phase Ar+$N_2$ CRAD with $N_2$ content of 0.11% in the liquid and 0.30% in the gas phase, the maximum gain amounts at 1000 (Fig. 11) and 900 (Fig. 12) for the 2THGEM and 2THGEM/GEM/PCB multiplier respectively, which should be compared to those of two-phase Ar CRADS, namely to 1200 (Fig. 5) and 4500 (Fig. 9) respectively.

There is another serious drawback of the Ar+$N_2$ system, studied in detail in [33]: at higher ionization densities and larger $N_2$ content, the ionization yield from a track in the liquid is considerably reduced compared to that in pure Ar. In this work, we confirmed this effect for signals induced by 60 keV X-rays, by comparing the amplitude distributions in the two-phase CRAD in Ar+$N_2$ (0.55% in the liquid, 1.5% in the gas) to that of pure Ar operated at similar gain (see Fig. 6): for the former the 60 keV peak amplitude is reduced by a factor of 2 compared to that of the latter. Similar reduction factor was observed for the smaller $N_2$ content studied in this work.

Thus we may conclude that the two-phase CRADs operated in Ar+$N_2$, though having faster signals due to conversion of the slow electron emission component to that of fast emission, do not offer the advantage in terms of the maximum gain compared to that of pure Ar. Moreover, the presence of $N_2$ additives results in a reduced ionization yield from a track for higher ionization density, presumably due to enhanced non-ionizing energy relaxation via $N_2$ molecules. These drawbacks make difficult the application of two-phase Ar+N$_2$ CRADs in low-threshold rare-event experiments.

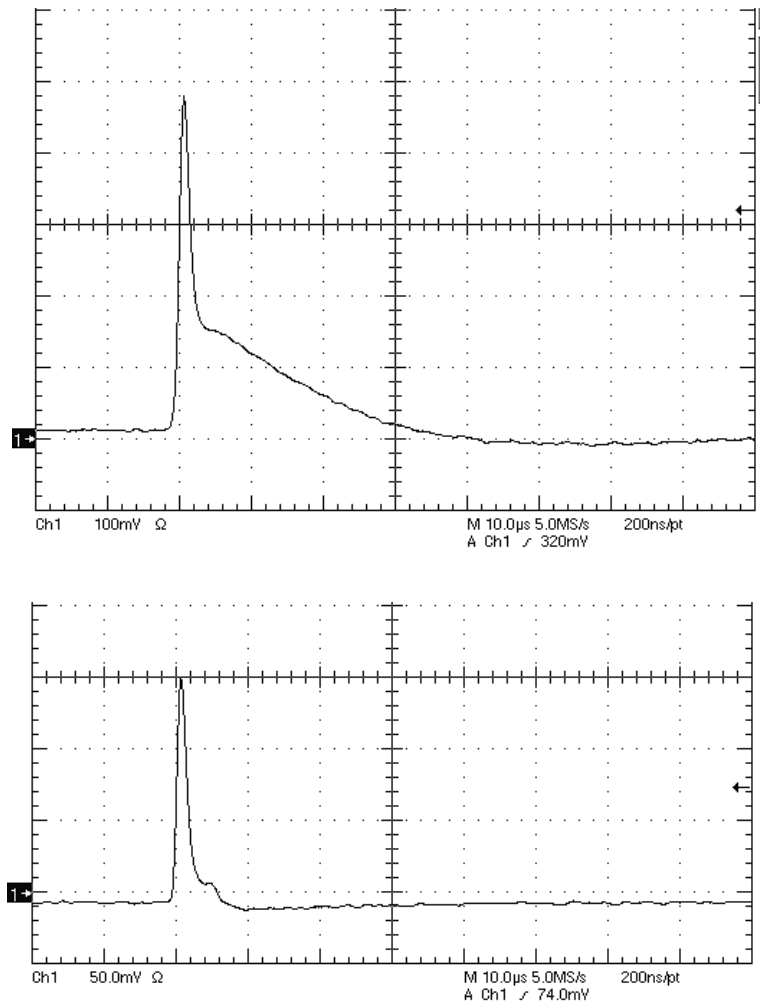

Fig. 10. Typical averaged anode signals in the two-phase CRADs with the 2THGEM/GEM/PCB multiplier at an electric field in the liquid at the liquid-gas interface of 1.6 kV/cm. Top: in Ar at an overall gain of 1700 and 2THGEM gain of 460; the fast and slow components are distinctly seen, the slow component being dominant. Bottom: in Ar+N$_2$ (0.11% in the liquid, 0.30% in the gas) at an overall gain of 680 and 2THGEM gain of 810, the slow component being almost fully converted to the fast one. The time scale is 10 μs/div; the vertical scale is 100 mV/div (top) and 50 mV/div (bottom). The charge amplifier shaping time is 0.5 μs.

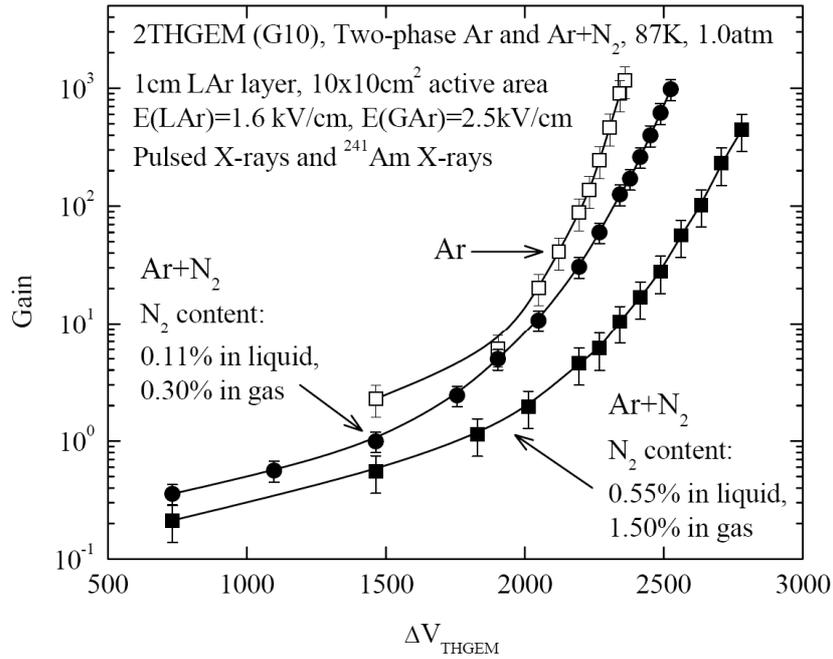

Fig. 11. Gain-voltage characteristics in the two-phase 10×10 cm² Ar and Ar+N₂ CRADs with the 2THGEM multiplier and 1 cm thick liquid layer. Here the maximum gains were limited by discharges. The THGEM geometrical parameters are t/p/d/h=0.4/0.9/0.5/0.1 mm. The electric fields and N₂ contents in the liquid and gas phases are indicated in the figure.

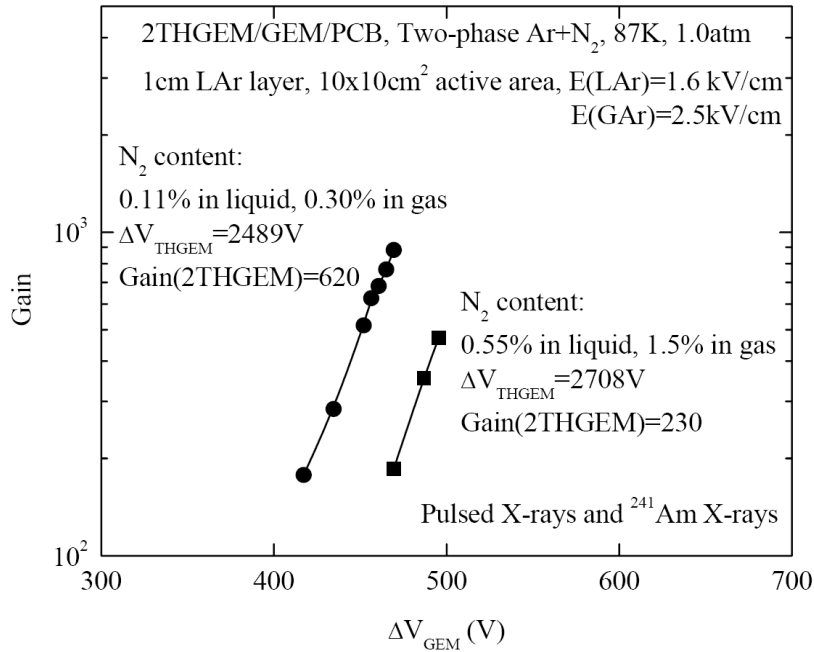

Fig. 12. Gain-voltage characteristics in the 10×10 cm² two-phase Ar+N₂ CRAD with the 2THGEM/GEM/PCB multiplier and 1 cm thick liquid layer. Shown is the overall multiplier gain as a function of the voltage across the GEM, at a fixed voltage across each THGEM, i.e. at a given 2THGEM gain, at two different N₂ contents. Here the maximum gains were limited by discharges. The electric fields and N₂ contents in the liquid and gas phases are indicated in the figure.

## 5. Two-phase CRAD with polyimide 2THGEM multiplier in Ar

The performance of a two-phase Ar CRAD with the polyimide 2THGEM multiplier was not satisfactory at this stage (see Fig. 13). A rather low maximum gain, ~ 30, and a tendency to enhanced operation voltages were observed. These were accompanied by gain instabilities appearing in the form of large gain fluctuations. These instabilities were similar to those observed earlier in our laboratory for a THGEM with resistive electrodes (RETHGEM) operated in two-phase Ar [7], confirming the hypothesis that certain multiplier types have difficulties of operation in saturated vapour [1]. Note that the geometrical parameters of the polyimide THGEMs and RETHGEMs were somewhat different from the regular THGEMs; e.g. they had smaller thickness (0.25 mm) and hole diameters (0.3 mm) as shown in Table 1. The most rational explanation of these instabilities is the effect of vapour condensation within the THGEM holes that prevents proper electron multiplication. The latter hypothesis was confirmed by the observation demonstrated in Fig. 13, that the polyimide 2THGEM multiplier was successfully operated in low-temperature gaseous Ar at a gas density corresponding to that of saturated vapor in the two-phase mode, with gains reaching 2000.

However another trouble has been revealed in this case. The polyimide THGEM turned out to have low resistance to discharges: a fatal discharge occurred at the maximum voltage (indicated in Fig. 13), causing total damage of this electrode.

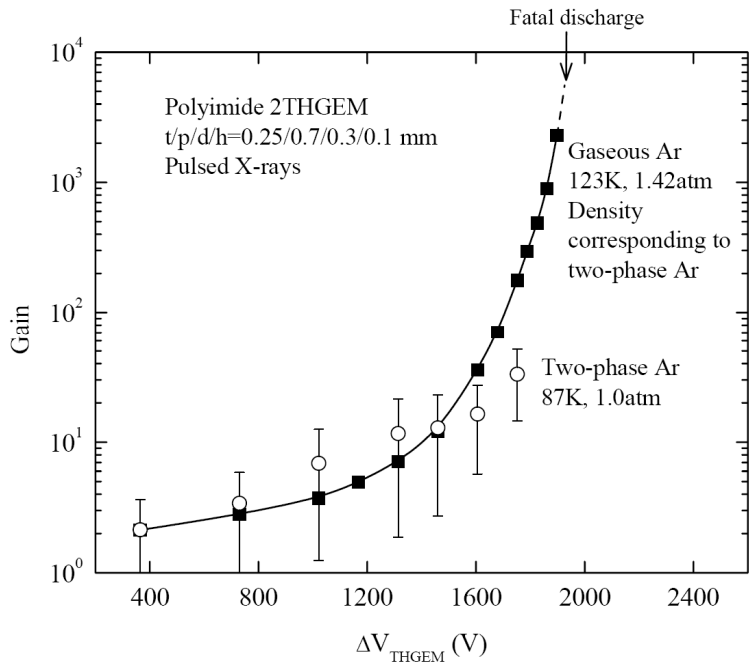

Fig. 13. Gain-voltage characteristics in the two-phase Ar CRAD with the polyimide 2THGEM multiplier and 1 cm thick liquid layer; the large gain fluctuations observed here are indicated by the error bars. The gain-voltage characteristics of the polyimide 2THGEM multiplier in gaseous Ar at cryogenic temperature is also shown, at gas density corresponding to that of saturated vapour in the two-phase mode; here a fatal discharge occurred at the maximum voltage (indicated in the figure). The operation conditions and THGEM geometrical parameters are indicated in the figure.

It should be remarked that the present polyimide THGEM design was not optimized in terms of resistance to discharges. In particular, the THGEM surface was fully metallized on one side; also the holes for the supporting structure were drilled directly through this metallization, as seen in Fig. 3; it could have resulted in edge discharges.

In conclusion, the two-phase Ar CRAD with the present polyimide 2THGEM multiplier showed poor performance in the two-phase mode and low resistance to discharges, perhaps due to poor choice of the THGEM design.

| Two-phase medium | Multiplier type | Active area | Typical maximum gain | Reference |
|---|---|---|---|---|
| Ar | 3GEM | 2.8×2.8 cm$^2$ | (5-10)×10$^3$ | [5],[6] |
| Ar | 2THGEM | 2.5×2.5 cm$^2$ | 3000 | [7] |
| Ar | 1THGEM | 2.5×2.5 cm$^2$ | >200 | [7] |
| Ar | 1THGEM | 4.5 cm in diameter | 300 | [15] |
| Ar | 1THGEM/PCB | 10×10 cm$^2$ | 80 | [8] |
| Ar | 2THGEM | 10×10 cm$^2$ | 1200 | This work |
| Ar | 1THGEM | 10×10 cm$^2$ | 300 | This work |
| Ar | 2THGEM/GEM/PCB | 10×10 cm$^2$ | 4500 | This work |
| Ar+N$_2$(0.1%) | 2THGEM | 10×10 cm$^2$ | 1000 | This work |
| Ar+N$_2$(0.1%) | 2THGEM/GEM/PCB | 10×10 cm$^2$ | 900 | This work |
| Ar | 2GEM(polyimide) | 5 cm in diameter | 30 | This work |

Table 2. Summary of maximum charge gains reached in two-phase Ar and Ar+N$_2$ CRADs. All THGEMs presented in the table were made of G10, except of that of the last row which was made of polyimide.

## 6. Conclusions

The present work continues a series of studies of GEM and THGEM multipliers in two-phase noble gases; we therefore summarized in Table 2 the available data on the maximum charge gain accumulated with these hole-multipliers in two-phase Cryogenic Avalanche Detectors (CRADs) operated in Ar and Ar+N$_2$.

We confirmed here the performance of two-phase CRADs in Ar with the double-THGEM multiplier, having an active area, of 10×10 cm$^2$, and a litre-scale liquid phase volume. Relatively high charge gains, of about 1000, were reached.

Higher gains, of about 5000, were attained in two-phase Ar CRADs with a hybrid triple-stage multiplier, comprising of a double-THGEM followed by a GEM (2THGEM/GEM/PCB). In this configuration, a patterned PCB readout anode has a practical importance in position-sensitive charge detectors.

The performance studies of two-phase CRADs operated in Ar doped with $N_2$ (0.1-0.6%) yielded faster signals due to conversion of the slow electron emission component to that of fast emission; however, such conditions resulted in a reduced ionization yield from higher ionization-density tracks and did not offer higher maximum gains compared to the operation in pure Ar. These would make difficult the application of two-phase Ar+$N_2$ CRADs in dark matter search and coherent neutrino scattering experiments.

The two-phase Ar CRAD operated with a polyimide double-THGEM multiplier, presented here rather poor performance, namely unstable operation in the two-phase mode (compared to high gain reached at low temperature) and low resistance to discharges; it might have resulted from a poor design of this first polyimide THGEM electrode. Further studies on improved electrodes are planned.

It should be noted that the maximum gain values obtained so far in two-phase Ar CRADs are still not sufficient for an operation with charge readout at ultimate single-electron counting sensitivity. Nevertheless, the present results should be valuable for a number of other CRAD applications. Such detectors could reach the desired sensitivities using GAPD- or LAAPD-based optical readout, as briefly discussed in the text and reviewed in [1].

## 7. Acknowledgements

We are grateful to A. Chegodaev and R. Snopkov for technical support. This work was supported in part by the Ministry of Education and Science of Russian Federation and Grants of the Government of Russian Federation (11.G34.31.0047), the Russian Foundation for Basic Research (12-02-91509-CERN_a and 12-02-12133-ofi_m) and the Israel Science Foundation (477/10).